\documentclass[aps,pra,twocolumn,superscriptaddress,
floatfix]{revtex4-1}
\usepackage{amsmath}
\usepackage[next]{inputenc}
\usepackage[dvips]{epsfig}
\usepackage{bbm,bm,bbold}
\usepackage{booktabs}
\usepackage{svg} 
\usepackage{multirow}
\usepackage{hhline}
\usepackage{comment}
\usepackage{float}
\usepackage{enumitem} 

\usepackage{amsmath,amsfonts,amssymb,amsthm}
\usepackage{color}
\usepackage{graphicx,latexsym}

\usepackage[colorlinks=true,urlcolor=blue,citecolor=blue,linkcolor=blue]{hyperref}

\usepackage{lipsum}

\usepackage{nameref}
\usepackage{varioref}
\usepackage{cleveref}

\usepackage{natbib}

\begin{document}
\renewcommand{\vec}{\mathbf}
\renewcommand{\Re}{\mathop{\mathrm{Re}}\nolimits}
\renewcommand{\Im}{\mathop{\mathrm{Im}}\nolimits}
\newcommand\scalemath[2]{\scalebox{#1}{\mbox{\ensuremath{\displaystyle #2}}}}

\title{Giant resonant skew scattering of plasma waves in a two-dimensional electron gas}

\author{Cooper Finnigan}
\email{cooper.finnigan@monash.edu}
\affiliation{School of Physics and Astronomy, Monash University, Victoria 3800, Australia}

\author{Dmitry K. Efimkin}
\email{dmitry.efimkin@monash.edu}
\affiliation{School of Physics and Astronomy, Monash University, Victoria 3800, Australia}
\affiliation{ARC Centre of Excellence in Future Low-Energy Electronics Technologies, Monash University, Victoria 3800, Australia}

\begin{abstract}
Electron skew scattering by impurities is one of the major mechanisms behind the anomalous Hall effect in ferromagnetic nanostructures. It is particularly strong at the surface of topological insulators where electron dynamics is governed by the spin-1/2 Dirac equation. Motivated by recently discovered mappings between hydrodynamics and the spin-1 Dirac equation, we consider the scattering of plasma waves  --- propagating charge density oscillations --- excited in graphene off a non-uniform magnetic field created by an adjacent circular micromagnet. The calculated scattering amplitude not only exhibits a giant asymmetry, or skewness, but is resonantly enhanced if the frequency of the incoming wave matches the frequency of the chiral trapped mode circulating the micromagnet in only one direction. Furthermore, if the frequency of the incoming plasma wave is a few times larger than the Larmor frequency, the angular distribution of its forward scattering is almost indistinguishable from that of a Dirac electron at the surface of a topological insulator scattering off a magnetic impurity. The micrometer scale of the proposed setup enables direct investigations of individual skew scattering events previously inaccessible in electronic systems.
\end{abstract}

\date{\today}
\maketitle
\textcolor{blue}{\emph{Introduction}}: Spin--orbit interactions give rise to a plethora of phenomena in solids and play a central role in magnetic-field-free spintronic devices. These interactions have a relativistic origin and are weak in conventional semiconductor nanostructures; yet, the interactions dominate the kinetic energy of Dirac electronic states at the surface of topological insulators, e.g., $\hbox{Bi}_2 \hbox{Se}_3$ and $\hbox{Bi}_2 \hbox{Te}_3$~\cite{TiReview1,TiReview2}. The emergent spin--momentum locking leads to coupled charge and spin transport~\cite{TITransport1}, strong inverse spin-galvanic effects~\cite{TIInverseGalvanic}, and a rich interplay with topological magnetic defects~\cite{TIDeffects1,TIDeffects2,TIDeffects3,TIDeffects4} and spin waves~\cite{TISpinWaves1,TISpinWaves2,TISpinWaves3} in the presence of long-range magnetic ordering~\cite{TIMagneticProximity}. These effects can be exploited in low-energy electronic and spintronic applications~\cite{SpintronicsReview}. 

Asymmetric -- or skew -- impurity scattering is another prominent effect due to the interplay between spin--orbit interactions and magnetism, resulting in the anomalous Hall effect (AHE)~\cite{AHEReview1}. The average scattering angle for Dirac electrons has been found to be an order of magnitude larger than the typical angle observed in conventional semiconductor systems, and the enhanced AHE has already attracted much attention~\cite{AHETI1,AHETI1_v2,AHETI2,AHETIexp1,AHETIReview}. However, there are other contributions, Berry phase-induced intrinsic and side-jumps mechanisms, and their separation is not straightforward. Moreover, the AHE is usually measured via macroscopic current--voltage probes, which average over the macroscopic number of electronic collisions. It would be highly desirable to develop a platform in which the skew scattering of Dirac electrons is tunable and can be directly probed.

\begin{figure}[b]
    \centering
    \vspace{-0.4cm}
    \includegraphics[width=0.9\columnwidth]{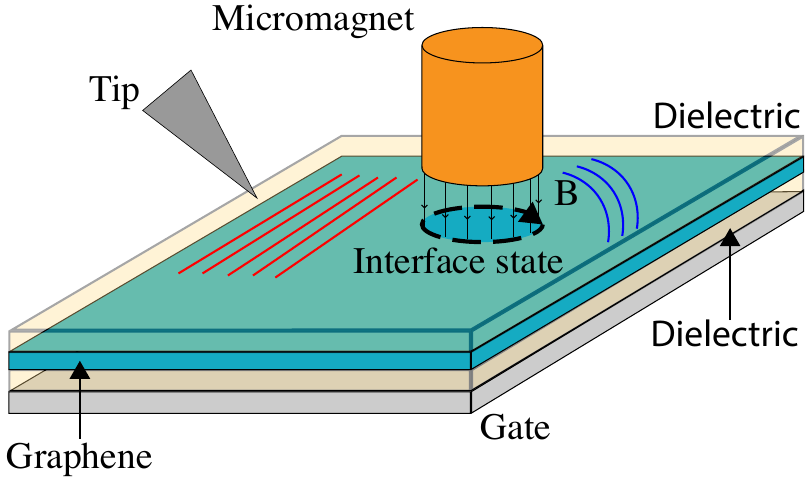}
    \vspace{-0.1cm}
    \caption{Schematic illustration of the considered setup. Plasma waves are excited in graphene by a near-field source and scatter off a non-uniform magnetic field created by an adjacent circular micromagnet. The calculated scattering amplitude has a high skewness and resonant behavior due to a single chiral interface mode (blue arrow line) circulating the micromagnet.}
    \label{Sketch}
    \vspace{-0.3cm}
\end{figure}

Recently, it has been realized that plasma waves (or plasmons) --- propagating charge density oscillations --- supported by a two-dimensional electron gas can be seen as distinct relatives of Dirac electrons~\cite{MagnetoPlasmons1}. The coupled hydrodynamic and Poisson equations describing their long-wavelength behavior can be reformulated as the relativistic-like pseudospin-1 Dirac equation. This mapping uncovered the topological stability of edge magnetoplasma waves~\cite{Fetter1} and established the path toward their topological engineering~\cite{InterfaceMPFinnigan, InterfaceMP1,InterfaceMP2,InterfaceMP4}. These findings naturally raise the question of whether the unexpected Dirac nature and resulting pseudospin--wavevector locking favor their skew scattering and whether such observations can provide helpful insights into the skew scattering of Dirac electrons. 

In the present Letter, we consider plasma wave scattering in the setup sketched in Fig.~\ref{Sketch}. Propagating plasma waves in gated graphene, which is a unique, tunable, low-loss plasmonic material~\cite{GraphenePlasmonicsReview1,GraphenePlasmonics1,GraphenePlasmonics2}, are excited via a near-field source (e.g, a capacitive
injector or atomic force microscope tip). The waves scatter off a non-uniform magnetic field created by an adjacent circular micromagnet. The calculated angular scattering distribution has a giant skewness, characterized by an average scattering angle exceeding $\pi/4$. The scattering is further resonantly enhanced if the frequency of the incoming wave matches the frequency of the chiral trapped mode circulating the micromagnet in only one direction.  Furthermore, if the frequency of the incoming plasma wave is a few times larger than the Larmor frequency, the angular distribution of its forward scattering is almost indistinguishable from that of a Dirac electron at the surface of a topological insulator scattering off a magnetic impurity.   
The micrometer scale of the proposed setup can enable direct probing of skew scattering events previously inaccessible in electronic systems. 

\textcolor{blue}{\emph{Hydrodynamic framework}}: 
The long-wavelength behavior of plasma waves in graphene can be described by classical hydrodynamic equations~\cite{Fetter1}. The linearized  equations for electron density $\rho(\mathbf{r},t)$ and electric current $\mathbf{j}(\mathbf{r},t)$ can be presented as
\begin{equation}
\label{HydroEq}
\begin{gathered}
\partial_t \rho(\mathbf{r}, t)+\nabla \cdot \mathbf{j}(\mathbf{r}, t)=0, \\
\partial_t \mathbf{j}(\mathbf{r}, t)=-\frac{n e^2}{m} \mathbf{\nabla} \phi(\mathbf{r}, t)+\frac{e}{m c}[\mathbf{j}(\mathbf{r}, t) \times \mathbf{B}(\mathbf{r})].
\end{gathered}
\end{equation}
Here, $m$ is the cyclotron mass for electrons, $n$ is their equilibrium concentration, and $\mathbf{B}(\mathbf{r})$ is the external magnetic field perpendicular to the graphene sheet. If the graphene is gated, the scalar potential $\phi(\mathbf{r},t)$ created by the charge density fluctuations is overscreened and can be approximated as $\phi(\mathbf{r},t)=\rho(\mathbf{r},t)/C$. The capacitance per unit area $C=\kappa/4\pi d$ is determined by the distance to the gate $d$ and the dielectric constant of the separating medium $\kappa$. 

If the external magnetic field is uniform, the equations support plasma waves with the gapped dispersion relation $\Omega_\vec{k}= \sqrt{(u k)^2+\omega_\mathrm{c}^2}$, where $u=\sqrt{4\pi n e^2 d /m \kappa}$ is the velocity of plasma waves in the absence of a magnetic field and $\omega_\mathrm{c}=e B /m c$ is the Larmor frequency. The dispersion relation $\Omega_{\vec{k}}$ has a relativistic-like appearance, and it has been recently found that this feature is not a coincidence~\cite{MagnetoPlasmons1}. The hydrodynamic equations can be rewritten as $i \partial_t \psi(\vec{r},t)=\hat{H}(\hat{\vec{k}}) \psi(\vec{r},t)$,  where
$\psi (\mathbf{r},t) =\left[j_{+}(\mathbf{r},t), u \rho (\mathbf{r},t),j_{-}(\mathbf{r},t)\right]$ with  $j_{\pm}(\mathbf{r},t)=\left[j_{x}(\mathbf{r},t) \pm i j_{y}(\mathbf{r},t) \right]/\sqrt{2}$ and $\hat{\vec{k}}=-i \mathbf{\nabla}$. The Hermitian matrix $\hat{H}(\hat{\vec{k}})$ is given by 
\begin{equation}
\hat{H}(\hat{\vec{k}})=\left(\begin{array}{ccc}
\omega_\mathrm{c}(\vec{r}) & \frac{u (\hat{k}_x-i \hat{k}_y )}{\sqrt{2}} & 0 \\
\frac{u (\hat{k}_x+i \hat{k}_y )}{\sqrt{2}}  & 0 & \frac{u (\hat{k}_x-i \hat{k}_y )}{\sqrt{2}}  \\
0 & \frac{u (\hat{k}_x+i \hat{k}_y )}{\sqrt{2}}  & -\omega_\mathrm{c}(\vec{r})
\end{array}\right), \label{eq 2.1}
\end{equation}
and plays the role of the Hamiltonian. This matrix can be presented as $\hat{H}(\vec{k})=u \hat{k}_x \hat{S}_x + u \hat{k}_y \hat{S}_y + \omega_c(\vec{r}) \hat{S}_z$, where the components of $\hat{\vec{S}}$  are the spin-1 generalization of the Pauli spin matrices. Interestingly, these reformulated equations are equivalent to the relativistic-like pseudospin-1 Dirac theory, and the Larmor frequency plays the role of the Dirac mass. The classical nature of the problem is reflected by the presence of particle--hole symmetry~\footnote{The symmetry works as $C H(\mathbf{q}) C^{-1}=-H(-\mathbf{q})$. The explicit expression for $C$ is given by 
$$C=
\begin{pmatrix}
0 & 0 & \mathcal{K} \\
0 & \mathcal{K} & 0 \\
\mathcal{K} & 0 & 0 \\
\end{pmatrix},
$$
where $\mathcal{K}$ is the complex conjugation operator. Only the positive frequency states correspond to physical modes supported by two-dimensional electron gas. }, which guarantees that any observables [e.g., $\vec{j}(\vec{r,t})$, $\rho(\vec{r},t)$, etc.] are real numbers. 

The spin-1 reformulation uncovers an emergent pseudospin--wavevector locking that is intricately connected with plasma wave polarization (relative amplitude and phase of propagating charge and current density oscillations). The pseudospin is directed along the vector $\vec{h}(\vec{k})$, which parameterizes the effective Hamiltonian as $\hat{H}(\vec{k})= \vec{h}(\vec{k})\cdot \hat{\vec{S}}$. The pseudospin has a meron-like texture across the reciprocal space and reduces to a vortex-like texture if the magnetic field vanishes. In the latter case, the corresponding spinor wave function simplifies as $|\vec{k}\rangle=\{e^{-i \phi_{\vec{k}}}, \sqrt{2}, e^{i \phi_{\vec{k}}} \}/2$ and describes the incident plasma wave in the considered scattering setup. 

\textcolor{blue}{\emph{Scattering setup}}: In the considered setup sketched in Fig.~\ref{Sketch}, the micromagnet plays not only the role of a target, but provides additional gating, affecting plasma waves in a twofold manner. First, the micromagnet creates a magnetic field that can be accurately approximated as nonzero and uniform only below the gate, i.e., $\omega_{\mathrm{c}}(\vec{r})=\omega_\mathrm{g} \Theta (r_\mathrm{g} -r)$~\footnote{This step-like profile corresponds to a ferromagnetic column with an infinite height. As argued in SM, our main results are valid even if the column height is comparable with its radius.}. This effect is the primary focus of this Letter.  Second, additional screening by free gate electrons modifies the plasma wave velocity as $u_\mathrm{g}=u\sqrt{n_\mathrm{g} d_\mathrm{g}/ n (d+d_\mathrm{g})}$. Here $d_\mathrm{g}$ is the distance to the micromagnet, and $n_\mathrm{g}$ is the electron concentration below it. As discussed in the Supplemental Material (SM), the velocity mismatch affects the plasma wave scattering only at much larger frequencies compared to the range discussed below. Besides, the second effect can be avoided for the insulating micromagnets (e.g.  $\hbox{Fe}_3\hbox{O}_4$).   


Another advantage of the spin-1 reformulation is that it offers scattering theories developed for electronic systems~\cite{AHEReview1} (for different approaches to plasma wave scattering, see ~\cite{PlasmonScattering1,PlasmonScattering2,PlasmonScattering3,PlasmonScattering4,PlasmonScatteringNearGate,Costaspaper1,GyrotropicScattering}). Furthermore, we can establish a connection with the scattering problem for Dirac electrons at the surface of a topological insulator interacting with deposited magnetic impurities with out-of-surface magnetic moments ~\cite{TIMagneticImpurityScattering} (or magnetic textures ~\cite{TISkewSkyrmion1,TISkewSkyrmion2}). The corresponding exchange field influencing the electrons has been described as a disk-shaped Dirac mass profile~\cite{TIMagneticImpurityScattering}. The scattering strength in both spin-1 and spin-1/2 Dirac models can be described by the same dimensionless parameter $\alpha= r_\mathrm{g} \omega_{\mathrm{g}} / u$, and we compare these predictions below. 

\textcolor{blue}{\emph{Scattering theories}}:  In the weak-scattering regime $\alpha\ll 1$, the problem can be approached via perturbation theory. The scattering amplitude $f_{\vec{k}'\vec{k}}$ between the plasma wave states $|\vec{k}\rangle$ and $|\vec{k}'\rangle$ is given by the matrix element of the T-matrix as 
\begin{equation}
    f(\phi) = \sqrt{\frac{k}{2 \pi u^2}} \hat{T}_{\vec{k}' 
\vec{k}}(\Omega_{\vec{k}}). \label{eq 3.16}
\end{equation}
Skew scattering arises from the interference between the first- and second-order scattering processes. In the second-order Born approximation, the T-matrix is approximated as $\hat{T}_{\vec{k}' 
\vec{k}}(\omega)=\hat{V}_{\vec{k}' 
    \vec{k}} + \sum_{\vec{k}''}\hat{V}_{\vec{k}' 
    \vec{k}''}\hat{G}(\omega, \vec{k}'')\hat{V}_{\vec{k}'' 
    \vec{k}}$, where $G(\omega,\vec{k})=(\omega-\hat{H}(\vec{k})+i \epsilon)^{-1}$ is the Green function for the Hamiltonian $\hat{H}(\vec{k})$ describing plasma waves.

\onecolumngrid
\begin{center}
\begin{figure}[h]
\centering
\vspace{-0.1cm}
        \includegraphics[trim=3cm 2cm 2.cm 2cm, clip,width=1\textwidth, height=110mm]{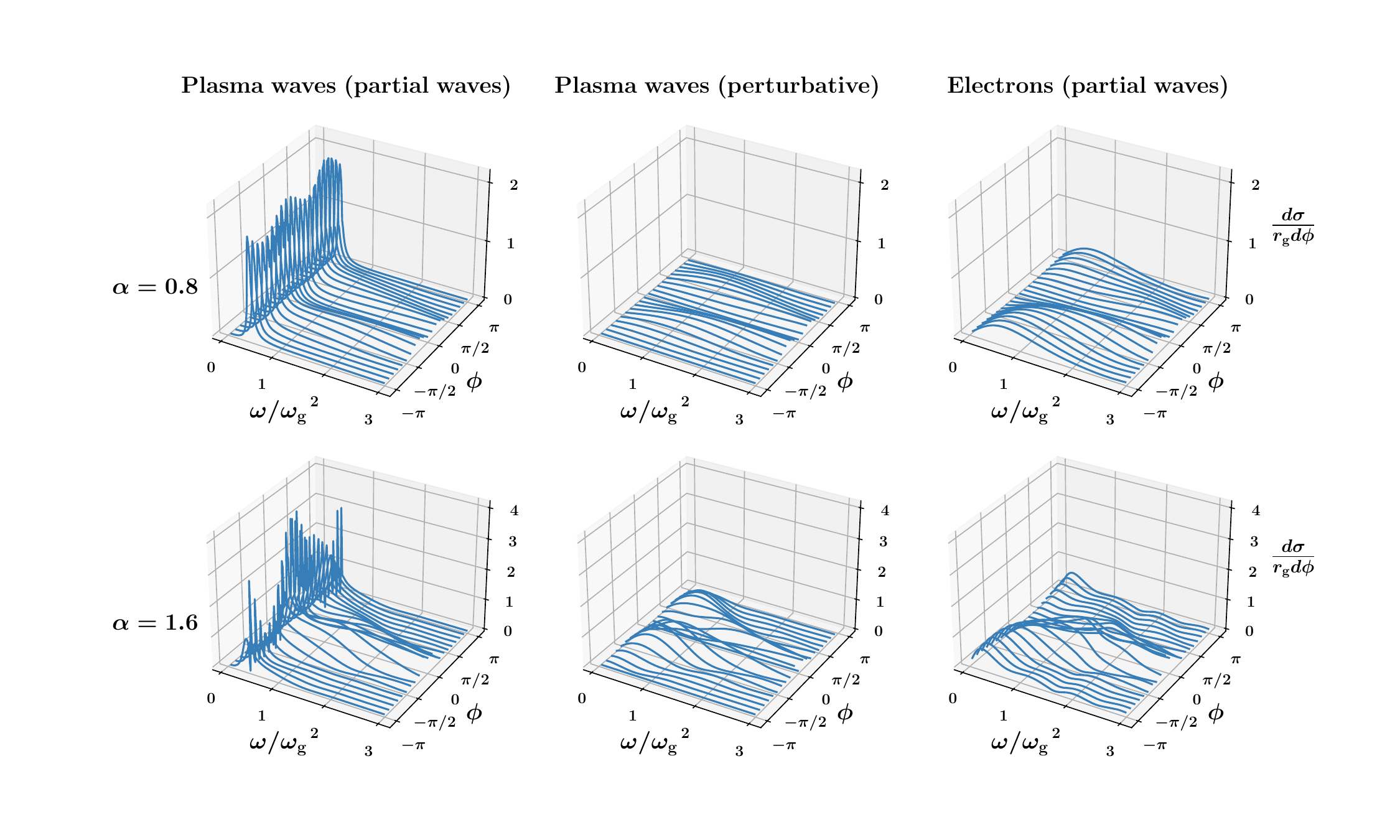}
        \caption{Angular and frequency dependence of the differential cross-section $d\sigma/d\phi$ for the scattering of plasma waves (left and middle columns) and electrons (right column). As stated in the top labels, these results were calculated by using either a partial wave or a second-order Born approximation with $\alpha=0.8$ (top row) or $\alpha=1.6$ (bottom row).}
        \label{Figure 2}
\vspace{-0.9cm}
\end{figure}
\end{center}
\twocolumngrid
As shown in the SM~\footnote{Supplementary material}, the differential cross-section $d\sigma / d\phi = |f(\phi)|^{2}$ is given by
\begin{equation}
\begin{split}
    \frac{d \sigma}{d \phi} \approx & \frac{k}{2 \pi u^{2}} \Big(|\hat{V}_{\mathbf{k}^{\prime} \mathbf{k}}|^{2} + \\ 
& 2\pi \sum_{\vec{k}''} \Im\big[\hat{V}_{\vec{k}' \vec{k}''} \hat{V}_{\vec{k}'' \vec{k}} \hat{V}_{\vec{k} \vec{k}'} \big]\delta [\Omega_{\vec{k}''} - \Omega_{\vec{k}}]\Big).
\end{split}
\label{PerturbativeCrossSection}
\end{equation}
The matrix element $\hat{V}_{\vec{k} \vec{k}'}=\omega_c(\vec{k}-\vec{k'}) \times M_{\vec{k}\vec{k}'}$ is shaped by the spatial profile of the magnetic field $\omega_\mathrm{c}(\vec{k})= 2 \pi r_{\mathrm{g}} \omega_{\mathrm{g}}J_{1}( k r_{\mathrm{g}})/k $ and by the matrix element of the spinor wave functions $M_{\vec{k}\vec{k}'}=\langle \mathbf{k} |\hat{S}_z | \mathbf{k}^{\prime} \rangle=i \sin(\phi_\vec{k}-\phi_\vec{k'})/2$. The latter is essential for skew scattering to emerge, confirming its intricate connection with the emergent pseudospin--wavevector locking. While the matrix element $\hat{V}_{\vec{k} \vec{k}'}$ vanishes at $\vec{k}=\vec{k}'$, the forward scattering is strongly suppressed only in the weak-scattering regime, as revealed by non-perturbative approaches.  
Because of the angular symmetry of the target, the scattering problem can be addressed in a non-perturbative manner via partial wave analysis. The asymptotic behavior for the wave function describing the plasma waves can be written as  
\begin{equation}
\psi(r, \phi) = \frac{1}{2}\left(\begin{array}{c}
1 \\
\sqrt{2} \\
1
\end{array}\right) e^{i k r \cos \phi}+\frac{f(\phi)}{2}\left(\begin{array}{c}
e^{-i \phi} \\
\sqrt{2} \\
e^{i \phi}
\end{array}\right) \frac{e^{i k r}}{\sqrt{ i r}}. \label{eq 3.12}
\end{equation}
The first term describes an incident and passed plane wave,
whereas the second term represents the scattered wave. The scattering amplitude $f(\phi)$ can be presented as  
\begin{equation}
f(\phi)=\sqrt{\frac{1}{2 \pi k}} \sum_m e^{i m \phi}\left[1-e^{-2 i \delta_m(k)}\right], \label{eq 3.13}
\end{equation}
where $\delta_{m}(k)$ is the phase shift for the partial wave labeled by the discrete orbital number $m$.
These phase shifts can be calculated using the radial equation for $\psi(r,\phi)$ supplemented with the boundary conditions ensuring electron charge conservation and continuity of electric potential at the interface where the magnetic field vanishes. These calculations are presented in the SM, whereas this Letter is focused on an analysis of the scattering observables. 

\textcolor{blue}{\emph{Scattering observables.}} The angular distribution of the plasma wave scattering is governed by the differential cross-section  $d\sigma/d\phi = |f(\phi)|^{2}$. The magnitude and skewness can be characterized by the total cross-section $\sigma$ and the average scattering angle $\langle \phi \rangle$, which are connected with the scattering amplitude as  
\begin{equation}
\begin{gathered}
\sigma= \int_{-\pi}^{\pi} |f(\phi)|^{2} d\phi, \quad \quad
        \langle \phi \rangle = \frac{1}{\sigma} \int_{-\pi}^{\pi} \phi |f(\phi)|^{2} \ d\phi \label{eq 3.18}
    \end{gathered}
\end{equation}
and depend only on the dimensionless frequency of the incoming plasma wave $\omega/\omega_{\mathrm{g}}$ and the controlling scattering parameter $\alpha$.

\begin{figure}[t]
    \centering
    \vspace{-0.2cm}
    \includegraphics[clip,trim={0cm 0.5cm 0cm 2cm}, width=\columnwidth]{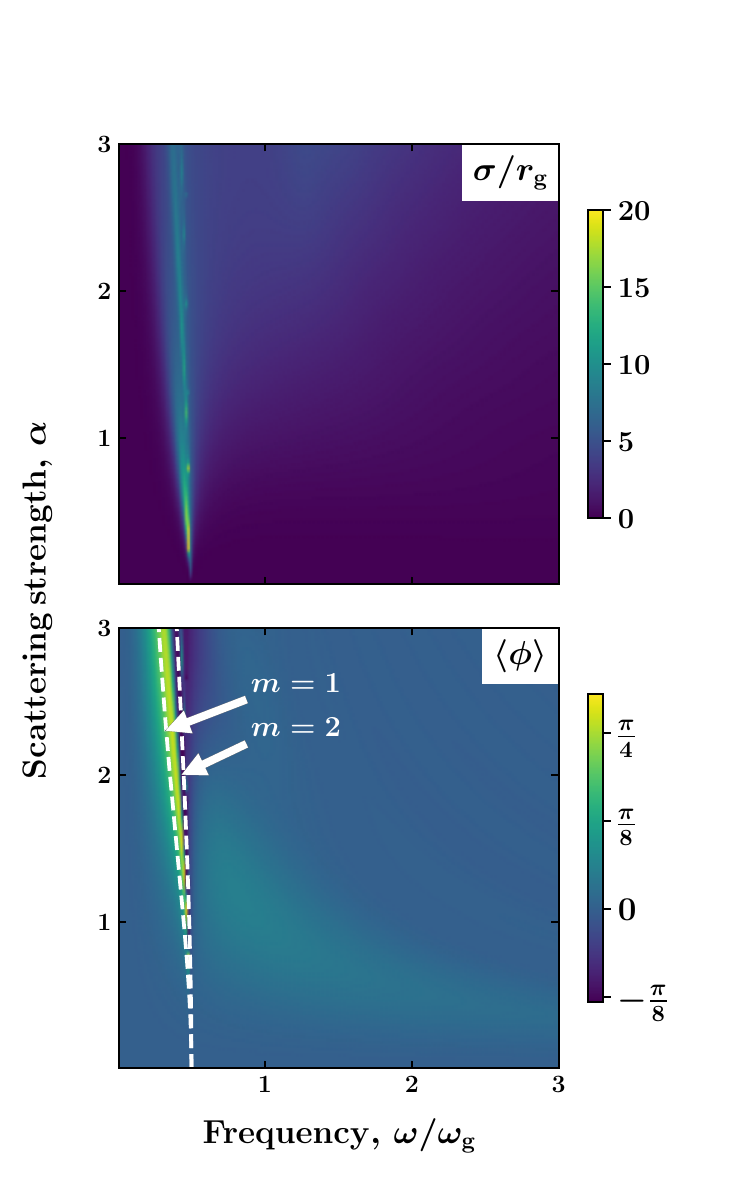}
    \caption{Scattering strength and frequency dependence of the total cross-section $\sigma$ (top) and average scattering angle $\langle \phi \rangle$ (bottom). The total cross-section is colored in a log scale, and the first two resonant harmonics $(m=1,2)$ are superimposed. }
    \label{Figure 3}
    \vspace{-0.5cm}
\end{figure}

Fig.~\ref{Figure 2} presents the angular distribution of the plasma wave scattering evaluated via the partial wave analysis (left column) and perturbation theory (middle column). The high-frequency results exhibit slightly visible oscillations (they become more prominent at smaller $\alpha$ or even higher frequencies), which approximately follow the squared Fourier transform of the cyclotron frequency profile $|\omega_\mathrm{c}(\vec{k}-\vec{k}')|^2$. Although the perturbation theory is expected to be reliable only in the weak-coupling regime $\alpha\ll 1$, it captures the high-frequency behavior reasonably well, even at $\alpha\sim 1$. In contrast, the perturbation theory completely fails to capture a prominent peak at low frequencies $\omega\lesssim \omega_{\mathrm{g}}/2$, which signals resonant plasma wave scattering. Before addressing its physical origin, we discuss the other scattering observables.  The dependence's of the total cross-section $\sigma$ and average scattering angle $\langle \phi\rangle$ on the frequency of the incoming plasma wave $\omega/\omega_{\mathrm{g}}$ and the dimensionless parameter $\alpha$ are presented in Fig. \ref{Figure 3}. The frequency dependence of the total cross-section $\sigma$ has a narrow peak at the resonance frequency. The average scattering angle achieves a maximum below the resonance, rapidly switches its sign across the resonance, and then has a prominent minimum. The magnitude of the extrema (approximately $\pi/4$ and $-\pi/8$, respectively) drastically exceeds the typical average angle for electron scattering in nanostructures. 

\textcolor{blue}{\emph{Resonant scattering and chiral interface modes}}. 
The presence of resonance is, perhaps, surprising. First, the magnetic field tends to open the local gap in plasma wave dispersion and, therefore, act repulsively. Second, the interface where the magnetic field vanishes does not trap classical electronic trajectories into skipping or snake orbits. Localized plasma waves, however, can be trapped at edges or different interfaces~\cite{Edge2DAnisotropic,EdgeDensity1,EdgeDensity2,EdgeDensity3, EdgeAnomalousHall, InterfaceMP1, InterfaceMP2, InterfaceMP3}. Here, we argue that the plasma waves are not only trapped at the interface where the magnetic field vanishes but also circulate the micromagnet in only one direction.

The resonant nature of the scattering can be tracked in the frequency behavior of the S-matrix $S_{m}(\omega)=e^{2 i \delta_{m}(\omega/\omega_{\mathrm{g}})}$ extended to the complex plane as $\omega=\omega'+i \omega''$. Resonant modes manifest as poles below the real axis. As shown in the SM, $S_{m}(\omega)$ has a single pole for partial waves with $m\geq 1$. Furthermore, the number of resonant modes $n_\mathrm{b}$ can be determined according to
\begin{equation}
n_{\mathrm{b}} = \frac{\delta_{m}(0) - \delta_{m}(\infty)}{\pi}, \quad n_{\mathrm{b}}=\begin{cases}  1, & m\geq 1, \\ 0, &  m<1, \end{cases}
\end{equation}
which constitutes a heuristic proof of Levinson's theorem~\cite{TextbookNewton1982}. Because the bound modes appear only for positive $m$, they correspond to chiral waves circulating the gate in only one direction.

 \begin{figure}[t]
    \centering
     \vspace{-0.5cm}
    \includegraphics[width=1\columnwidth, trim={0cm 0cm 0cm 0cm}]{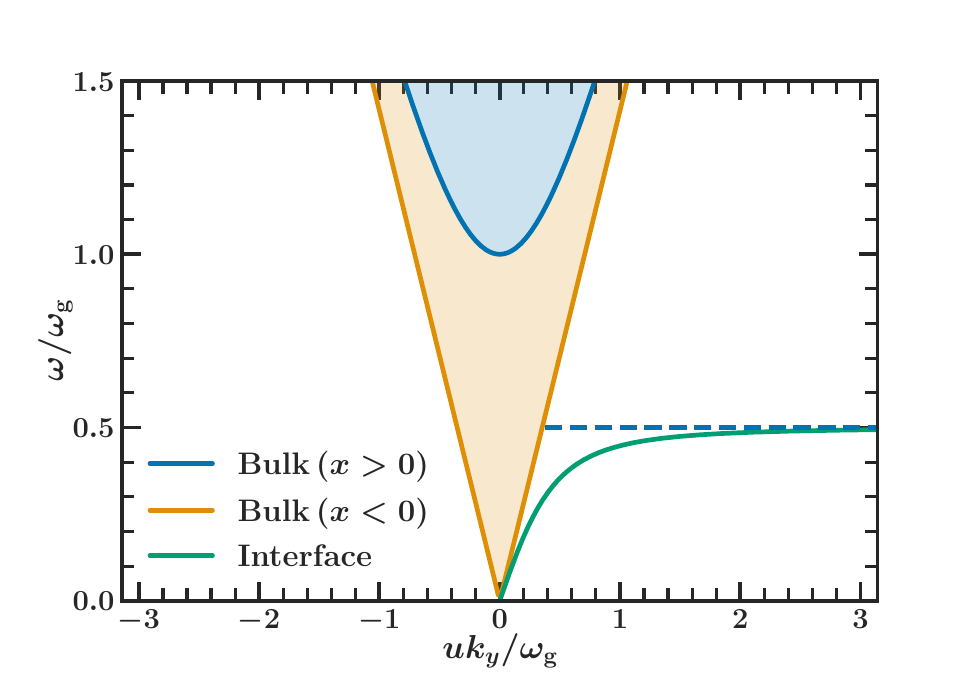}
    \caption{Dispersion of an interface plasma wave (green) localized at a magnetic field interface, i.e., $\omega_\mathrm{c}(x)=\omega_{\mathrm{g}} \Theta (x)$ and given by Eq.~(\ref{EqInterfaceDispersion}). Its frequency is below the continua of bulk states at both sides of the interface (orange and blue shaded areas). The edge mode propagates in only one direction and is, therefore, chiral.}
    \vspace{-0.4cm}
    \label{FigureEdge}
\end{figure}

To further confirm this picture, it is instructive to shift from the disc geometry to a half-plane geometry, i.e., $\omega_\mathrm{c}(x)=\omega_\mathrm{g}\Theta(x)$. This interface supports a single chiral trapped mode with the following dispersion relation: 
\begin{equation}
\label{EqInterfaceDispersion}
    \Omega_{\text{int}}(k_{y}) = \frac{u k_{y} \ \omega_{\mathrm{g}}}{\sqrt{4(u k_{y})^{2}+ \omega_{\mathrm{g}}^{2}}}.
\end{equation}
As presented in Fig. \ref{FigureEdge}, the frequency of this mode is below the continua of bulk modes at both sides of the interface (red and blue shaded areas) and saturates towards $\omega = \omega_{\mathrm{g}}/2$. In the disc geometry, the standing wave condition $2\pi k_m r_\mathrm{g} =2\pi m$ results in discretization of the trapped mode wavevector. Therefore, the resonant frequencies can be approximated as $\Omega_m=\Omega_{\text{int}}(k_m)$ and can be presented as $\Omega_m=\omega_{\mathrm{g}} /\sqrt {(\alpha/m)^2+4}$. Superimposed in Fig.~\ref{Figure 3}, these frequencies accurately indicate the position of the scattering resonance, confirming that it originates from the chiral trapped mode~\footnote{It is worth mentioning that the chiral trapped states are not topological and not immune to the back-scattering, but still have some protection. For instance, the counter-propagating modes trapped at the opposite sides of the magnetic field stripe (or placed nearby micromagnets) are reluctant to hybridize since they are well separated in the reciprocal space, i.e. $k_y>0$ and $k_y<0$}.  


\textcolor{blue}{\emph{Comparison with the scattering of Dirac electrons.}} The angular distribution for the scattering of Dirac spin-1/2 electrons at the same disk-like Dirac mass profile is also presented in Fig.~\ref{Figure 2} (right column). The electronic scattering is also enhanced at low frequencies, but the dependence is smooth and resonance-free, reflecting the absence of trapped electronic states. In contrast, the dependencies tend to match in the range $\omega_\mathrm{g} \lesssim \omega \lesssim 4 \omega_\mathrm{g}$, where the forward scattering ($|\phi| \lesssim\pi/4$) gradually becomes dominant. At larger frequencies, $\omega \gtrsim 4 \omega_\mathrm{g}$, the forward scattering for the two seemingly unrelated models becomes almost indistinguishable~\cite{SM}. Since the origin of this behavior is not apparent within partial wave formalism, it is instructive to return to the perturbative expression for the differential cross-section, Eq.~(\ref{PerturbativeCrossSection}). In dimensionless units, both considered problems are assumed to have the same dispersion relation and target profile. The difference between them is hidden in the matrix element of the spinor wave functions $M_{\vec{k}\vec{k}'}$. It depends only on the scattering angle $\phi=\phi_{\vec{k}}-\phi_{\vec{k}'}$ and takes the form $M_{\vec{k}\vec{k}'}=i \sin(\phi)/2$ for plasma waves and $M_{\vec{k}\vec{k}'}=i \sin(\phi/2)$ for Dirac electrons~\footnote{These arguments imply that the target represents the spatially dependent Dirac mass and does not extend the other models, i.e. $\hat{H}_\mathrm{V}=V(\vec{r})\hat{1}$. The latter target model is not eligible for plasma waves since it breaks the particle-hole symmetry dictated by the classical nature of the problem but describes overscreened Coulomb impurities in spin-1 electronic systems~\cite{SkewScattering1,KleinScattering2,KleinScattering}}. They have different angular profiles but remarkably match each other for $|\phi|\lesssim\pi/4$, which can easily be seen by expanding both matrix elements with $\phi$ as a small parameter. The characteristic scattering angle can be approximated as $\sin(|\phi|/2)=q/2k$ where $q\approx1/r_\mathrm{g}$ and $k=\omega/u$ are magnitudes of transferred and incoming wave vectors. Thus, plasma waves and Dirac electrons scatter in the same manner if the forward scattering is dominant $|\phi|\ll \pi/4$, i.e. $\omega\gg 4 u/ \pi r_\mathrm{g}=4 \omega_\mathrm{g}/\pi\alpha$. This criterion is consistent with explicit calculations for circular and elliptical targets~\cite{SM}.  

\textcolor{blue}{\emph{Discussion}}: Plasma waves supported by a two-dimensional electron gas have been reported in various physical systems, including a liquid helium surface ~\cite{HeSurface1,HeSurface2}, silicon inversion layers~\cite{InversionLayer1}, graphene~\cite{GraphenePlasmonics1,GraphenePlasmonics2,GraphenePlasmonicsReview1}, and other two-dimensional materials~\cite{TMDplasmon1,TMDplasmon2}. In the considered setup, the monolayer graphene provides a few advantages. First, plasma waves have been reported there for a wide range of operating frequencies, spanning from mid-infrared to the terahertz range. Second, the extremely long relaxation times of monolayer graphene favor micrometer-sized plasma wave propagation.  Third, the gated geometry (e.g., graphene-$\hbox{SiO}_2$-Si) is commonly used and permits electrical tuning of the plasma wave dispersion and cyclotron frequency. The latter is possible because the cyclotron mass $m$ for relativistic-like electrons with linear dispersion $\epsilon_{\vec{p}}=v p$ is doping-dependent $m= \sqrt{n \pi \hbar^2}/v$ and is not protected by Kohn's theorem~\cite{KohnTherem}. Regarding the micrometer-sized magnet, conventional magnets such as Ni, NiFe, and NiCr have been used in similar setups~\cite{Micromagnets1,Micromagnets2,Micromagnets3,Micromagnets4,Micromagnets5} and have been reported to produce magnetic fields up to a few dozen mT.         

For estimations, we use the following set of parameters: electron velocity $v\approx 10^{8} \;\hbox{cm}/\hbox{s}$, electron concentration $n\approx 5 \times 10^{11}\; \hbox{cm}^{-2}$, distance to the gate $d\approx12\;\hbox{nm}$, effective dielectric constant for $\hbox{SiO}_2$ spacer $\kappa\approx 4$, magnetic field $B_\mathrm{g}\approx 30 \; \hbox{mT}$, and micromagnet radius $r_\mathrm{g}\approx3 \; \mu \hbox{m}$. The smallness of the cyclotron mass $m\approx0.15 \; m_0$ enhances the magnetic field effect at plasma waves. Here, $m_0$ is the free electron mass. The other parameters of the model are $u\approx 1.3 \times 10^{8} \; \hbox{cm/s}$ and $\omega_\mathrm{g}\approx 0.4 \; \hbox{THz}$, resulting in moderate-strength scattering $\alpha=0.8$. The cyclotron frequency is in the infrared range, where plasma waves in graphene have been previously reported~\cite{GraphenePlasmonicsReview1}.

In recent years, it has been found that the spin-1 Dirac model describes a diverse family of platforms, including photonic metamaterials~\cite{Spin1Photonic1,Spin1Photonic2}, superconducting qutrit circuits~\cite{Spin1SC1}, optical lattices in cold atomic setups~\cite{T3Model1,T3Model2}, and two-dimensional electronic systems with the $\alpha$-$T_3$ lattices, including  $\hbox{SrTiO}_3$-$\hbox{SrIrO}_3$-$\hbox{SrTiO}_3$ trilayer heterostructures~\cite{Spin1ElectronicThreelayer}, $\hbox{so-MoS}_2$ allotropes~\cite{Spin1ElectronicMoS2}, and graphene-$\hbox{In}_2 \hbox{Te}_2$ bilayers~\cite{Spin1ElectronicGraphene}. The Dirac mass term corresponds to a staggered sublattice potential and can be induced via substrate engineering. Thus, the predicted giant and resonant skew scattering can be potentially engineered in these systems, resulting in enhanced and tunable AHE. 

The chiral interface plasma wave mode circulating the micromagnet results in non-reciprocal plasma wave propagation and energy transfer~\cite{NonReciprical}. They can be probed using setups with multiple injectors and antennas, which have been dubbed efficient to directly probe edge magnetoplasma waves graphene flakes~\cite{PlasmonsGraphene1}.

To conclude, we predict giant resonant skew scattering of plasma waves off a micromagnet. The micrometer scale of the proposed setup can enable direct probes of individual skew scattering events previously inaccessible in electronic systems. Furthermore, observing skew plasma wave scattering could provide insights into the scattering of spin-1/2 and spin-1 Dirac electron scattering in nanostructures. 


\textcolor{blue}{\emph{Acknowledgments}}: 
We acknowledge fruitful discussions 
with Dimi Culcer, Oleg Sushkov, Michael Fuhrer, Haoran Ren and Stefan Maier and support from the Australian Research Council Centre of Excellence in Future Low-Energy Electronics Technologies (CE170100039). 

\bibliographystyle{apsrev4-1}
\bibliography{ReferencesSkewMP}

\newpage

\renewcommand{\theequation}{S\arabic{equation}}

\setcounter{section}{0}
\setcounter{equation}{0}
\setcounter{figure}{0}
\setcounter{enumiv}{0}

\clearpage

\setcounter{page}{1}

\onecolumngrid
\begin{center}
	\textbf{\large Supplemental Material for \\
		Giant resonant skew scattering of plasma waves in graphene off micromagnet}
	\\
	Cooper Finnigan and Dmitry K. Efimkin
\end{center}
\twocolumngrid

\section{The spin-1 reformulation}
This section discusses the reformulation of the coupled hydrodynamic and Poisson equations as a spin-1 Dirac equation. The propagating charge density oscillations $\rho\left(\mathbf{r}, t\right)$ induce the potential $\phi(\mathbf{r}, t)$ which is given by 
\begin{equation}
\phi(\mathbf{r}, t)=\int d \mathbf{r}^{\prime} V\left(\mathbf{r}-\mathbf{r}^{\prime}\right) \rho\left(\mathbf{r}^{\prime}, t\right), \; \; V(\mathbf{k})=\frac{2 \pi}{k \kappa(\mathbf{k})}.
\end{equation}
The potential $V(\mathbf{q})$ incorporates the screening by the external media as $\kappa(\vec{k})=\kappa /\tanh(k d)$~\cite{Fetter1}. Here $d$ is the distance to the gate, and $\kappa$ is the dielectric constant of the spacer. It is instructive to 
perform the Fourier transform, introduce the spinor $\psi(\mathbf{k},\omega) = \left\{j^{+}(\mathbf{k}, \omega), j^0(\mathbf{k}, \omega), j^{-}(\mathbf{k}, \omega)\right\}$ with components $j^{\pm}(\mathbf{k},\omega) = \frac{1}{\sqrt{2}}\left[ j^{x}(\mathbf{k}, \omega) \pm i j^{y}(\mathbf{k},\omega) \right]$ and rescale the electronic density as 
$j^0(\mathbf{k}, \omega)=\omega_{\mathrm{p}}(\mathbf{k}) \rho(\mathbf{k}, \omega) / k$. Here $\omega_{p}(\mathbf{k})= \sqrt{2 \pi n e^2 k \tanh(k d) / m \kappa }$ is the dispersion relation for magnetic field-free plasmons, which interpolates from the linear to the square-root dependence. The resulting equations can be presented as an eigenvalue problem $\omega \psi(\mathbf{k}, \omega)=\hat{H}(\mathbf{k}) \psi(\mathbf{k}, \omega)$. The Hermitian $3\times3$ matrix can be interpreted as the effective Hamiltonian of the system and is given by 
\begin{equation}
\label{HamiltonianGeneric}
\hat{H}(\mathbf{k})=\left(\begin{array}{ccc}
\omega_{c} & \frac{\omega_{\mathrm{p}}(\mathbf{k}) e^{i \phi_{\mathbf{k}}}}{\sqrt{2}} & 0 \\
\frac{\omega_{\mathrm{p}}(\mathbf{k}) e^{-i \phi_{\mathrm{k}}}}{\sqrt{2}} & 0 & \frac{\omega_{\mathrm{p}}(\mathbf{k}) e^{i \phi_{\mathbf{k}}}}{\sqrt{2}} \\
0 & \frac{\omega_{\mathrm{p}}(\mathbf{k}) e^{-i \phi_{\mathbf{k}}}}{\sqrt{2}} & -\omega_{c}.
\end{array}\right).
\end{equation}
Its eigenvalues have the three branches 
\begin{equation}
\Omega^{\pm}(\mathbf{k})= \pm \Omega(\mathbf{k}) \quad \text { and } \quad \Omega^0(\mathbf{k})=0. \label{Suppeq 3.8.2}
\end{equation}
The positive frequency branch $\Omega(\mathbf{k})=\sqrt{\omega^2_{\mathrm{c}}+\omega_{\mathrm{p}}^2(\mathbf{k})}$ governs the dynamics for plasma waves, and is gapped by the Larmor frequency. The negative-frequency branch is connected to the
positive branch by the particle-hole symmetry transformation and is not dynamically independent. The zero-frequency branch is spurious, and its flatness is protected by the interplay of inversion  $\left[\Omega^{0}(\mathbf{k}) = \Omega^{0}( -\mathbf{k}) \right]$  and the particle-hole $\left[ \Omega^{0}(-\mathbf{k}) = - \Omega^{0}(\mathbf{k}) \right]$ symmetries. If we approximate the dispersion of plasma waves by its long-wavelength linear behavior  $\omega_\mathrm{p}=u k$, the Hamiltonian reduces to the spin-1 Dirac model presented as Eq.~(2) in the main text of the Letter.  


\vspace{-0.5cm}
\section{The scattering phase shifts } \label{secII}
This section presents the derivation of scattering phase shifts for plasma waves. Due to the azimuthal symmetry of the magnetic field, it is natural to use the polar coordinates $\mathbf{r}=(r,\phi)$ and present the wave function for each angular harmonics $m$ as
\begin{equation}
    \psi_{m}(r,\phi,\omega)= \begin{pmatrix}
        \psi^{+}(r,\omega) e^{i(m-1)\phi} \\
        \psi^{0}(r,\omega) e^{im\phi} \\
        \psi^{-}(r,\omega) e^{i(m+1)\phi}
    \end{pmatrix}. \label{eq 3.3}
\end{equation}
The top and the bottom components of the spinor can be presented as 
\begin{gather}
    \psi^{+}(r,\omega)= \frac{-i u}{\sqrt{2}\left[\omega - \omega_{c}(\mathbf{r})\right]}\left(\partial_{r} + \frac{m}{r} \right)\psi^{0}(r,\omega), \\
    \psi^{-}(r,\omega) = \frac{-i u}{\sqrt{2} \left[\omega + \omega_{c}(\mathbf{r}) \right]}\left(\partial_{r} - \frac{m}{r}\right)\psi^{0}(r,\omega),
\end{gather}
and its central component $\psi^{0}(r,\omega)$ satisfies the closed equation as 
\begin{equation}
    \kappa^{2}(\vec{r})\psi^{0}(r,\omega) - \left(\partial_r^2+\frac{1}{r} \partial_r-\frac{m^2}{r^2}\right)\psi^{0}(r,\omega)=0.
\end{equation}
Here we have introduced $\kappa (\vec{r}) = \sqrt{|\omega^{2}-\omega_{c}(\mathbf{r})^{2}|}/u$. It is instructive to consider the inner ($r<r_{\mathrm{g}}$) and outer ($r>r_{\mathrm{g}}$) regions separately. They will be labeled as I and II, respectively.  In the outer region, the wave function can be presented as
\begin{equation}
    \psi^{\mathrm{II}}_{m} = B_{m} \begin{pmatrix} -i J_{m-1}(kr) \\
J_{m}(kr) \\
i J_{m+1}(kr)
\end{pmatrix} + C_{m} \begin{pmatrix} 
-i Y_{m-1}(kr) \\
Y_{m}(kr) \\
i Y_{m+1}(kr)
\end{pmatrix}. \label{eq 3.10}
\end{equation}
Here $k= \omega / u$ is the wavevector magnitude for the incoming plasma wave; $J_{m}$ and $Y_{m}$ are the Bessel functions of the first and the second kind. The wave function Eq.~(\ref{eq 3.10}) has oscillatory behavior, and its asymptotics differs from the one for the free plasma wave (in the absence of the micromagnet) only by the phase shift $\delta_{m}=\arctan\left(C_{m}/B_{m} \right)$. In contrast, the behavior of the wave function in the inner region depends on the relation between the frequency of incoming plasma wave $\omega$ and the local Larmor frequency $\omega_{\mathrm{g}}$ induced by the ferromagnetic gate. If $\omega>\omega_\mathrm{g}$, the wave function is also oscillatory and is given by
\begin{equation}
  \psi_{m}^\mathrm{I}=  A_m \begin{pmatrix}
\frac{-i}{\sqrt{2}} \sqrt{\frac{\omega+\omega_{\mathrm{g}}}{\omega-\omega_{\mathrm{g}}}}J_{m-1}(\kappa r)  \\
J_{m}(\kappa r) \\
\frac{i}{\sqrt{2}}\sqrt{\frac{\omega-\omega_{\mathrm{g}}}{\omega+\omega_{\mathrm{g}}}}J_{m+1}(\kappa r)
\end{pmatrix}. \label{eq 3.9}
\end{equation}
If $\omega <\omega_{g}$, the wave function is evanescent and decays toward the center of ferromagnet as 
\begin{equation}
    \psi_{m}^{\mathrm{I}}= A_m \begin{pmatrix} 
\frac{i}{\sqrt{2}} \sqrt{\frac{\omega_{\mathrm{g}}+\omega}{\omega_{\mathrm{g}}-\omega}}I_{m-1}(\kappa r) \\
I_{m}(\kappa r) \\
\frac{-i}{\sqrt{2}}\sqrt{\frac{\omega_{\mathrm{g}}-\omega}{\omega_{\mathrm{g}}+\omega}}I_{m+1}(\kappa r)
\end{pmatrix}. \label{eq 3.8}
\end{equation}
Here $I_{m}$ is the modified Bessel function of the first kind. 

To coefficients $A_m$, $B_{m}$ and $C_{m}$ can be found using boundary conditions at the interface. First, the continuity of the scalar potential dictates
\begin{equation}
\begin{aligned}
\psi_{m}^{0, \mathrm{I}}(r_{\mathrm{g}}) & =\psi_{m}^{0, \mathrm{II}}(r_{\mathrm{g}}),
\end{aligned} \label{eq 3.11}
\end{equation}
Second, charge conservation dictates the radial component of electric current density to be continuous across the interface, which implies
\begin{equation}
\psi_{m}^{+, \mathrm{I} }(r_{\mathrm{g}})+\psi_{m}^{-, \mathrm{I}}(r_{\mathrm{g}})  =\psi_{m}^{+, \mathrm{II}}(r_{\mathrm{g}})+\psi_{m}^{-, \mathrm{II}}(r_{\mathrm{g}}),
\label{eq 3.12}
\end{equation}
As for the azimuthal component of the electric current density, it can be discontinuous across the interface. These two boundary conditions are sufficient (all three of them can be calculated if we add the normalization condition for the wave function) to calculate the ratios $B_m/A_m$ and $C_m/A_m$, which determine plasma waves phase shifts as $\delta_{m}=\arctan\left(C_{m}/B_{m} \right)$. The explicit calculations result in 
 \begin{widetext}
\begin{equation}
\begin{aligned}
& \left(\frac{B_m}{A_m}\right)_{\omega<\omega_\mathrm{g}}=\frac{Y_m(k r_{\mathrm{g}}) \left(\sqrt{\frac{\omega +\text{$\omega_{\mathrm{g}} $}}{\text{$\omega_{\mathrm{g}} $}-\omega}} I_{m-1}(\kappa r_{\mathrm{g}} )-\sqrt{\frac{\text{$\omega_{\mathrm{g}} $}-\omega }{\omega +\text{$\omega_{\mathrm{g}} $}}} I_{m+1}(\kappa r_{\mathrm{g}})\right)+(Y_{m-1}(k r_{\mathrm{g}})-Y_{m+1}(k r_{\mathrm{g}})) I_m(\kappa r_{\mathrm{g}} )}{(J_{m+1}(k r_{\mathrm{g}})-J_{m-1}(k r_{\mathrm{g}})) Y_m(k r_{\mathrm{g}})+J_m(k r_{\mathrm{g}}) (Y_{m-1}(k r_{\mathrm{g}})-Y_{m+1}(k r_{\mathrm{g}}))}, \\
& \left(\frac{C_m}{A_m}\right)_{\omega<\omega_\mathrm{g}}=\frac{J_m(k r_{\mathrm{g}}) \left(\sqrt{\frac{\omega +\text{$\omega_{\mathrm{g}} $}}{\text{$\omega_{\mathrm{g}} $}-\omega}} I_{m-1}( \kappa r_{\mathrm{g}})-\sqrt{\frac{\text{$\omega_{\mathrm{g}} $}-\omega }{\omega +\text{$\omega_{\mathrm{g}} $}}} I_{m+1}(\kappa r_{\mathrm{g}})\right)+(J_{m-1}(k r_{\mathrm{g}})-J_{m+1}(k r_{\mathrm{g}})) I_m(\kappa r_{\mathrm{g}})}{(J_{m-1}(k r_{\mathrm{g}})-J_{m+1}(k r_{\mathrm{g}})) Y_m(k r_{\mathrm{g}})+J_m(kr_{\mathrm{g}}) (Y_{m+1}(k r_{\mathrm{g}})-Y_{m-1}(k r_{\mathrm{g}}))},
\end{aligned}
\end{equation}
$\qquad \qquad \qquad \qquad \qquad \qquad \qquad  \qquad \qquad \qquad \qquad \qquad$ and
\begin{equation}
\begin{aligned}
& \left(\frac{B_m}{A_m}\right)_{\omega>\omega_\mathrm{g}}=\frac{-Y_m(k r_{\mathrm{g}}) \left(\sqrt{\frac{\omega +\text{$\omega_{\mathrm{g}} $}}{\omega -\text{$\omega_{\mathrm{g}} $}}} J_{m-1}( \kappa r_{\mathrm{g}})-\sqrt{\frac{\omega -\text{$\omega_{\mathrm{g}} $}}{\omega +\text{$\omega_{\mathrm{g}} $}}} J_{m+1}(\kappa r_{\mathrm{g}})\right)+(Y_{m-1}(k r_{\mathrm{g}})-Y_{m+1}(k r_{\mathrm{g}})) J_m( \kappa r_{\mathrm{g}})}{(J_{m+1}(k r_{\mathrm{g}})-J_{m-1}(k r_{\mathrm{g}})) Y_m(k r_{\mathrm{g}})+J_m(k r_{\mathrm{g}}) (Y_{m-1}(k r_{\mathrm{g}})-Y_{m+1}(k r_{\mathrm{g}}))}, \\
& \left(\frac{C_m}{A_m}\right)_{\omega>\omega_\mathrm{g}}=\frac{-J_{m}(k r_{\mathrm{g}})\left(\sqrt{\frac{\omega + \omega_{\mathrm{g}}}{\omega- \omega_{\mathrm{g}}}}J_{m-1}(\kappa r_{\mathrm{g}}) + \sqrt{\frac{\omega - \omega_{\mathrm{g}}}{\omega+\omega_{\mathrm{g}}}}J_{m+1}(\kappa r_{\mathrm{g}}) \right)+ \left(Y_{m-1}(k r_{\mathrm{g}})-Y_{m+1}(k r_{\mathrm{g}})\right)J_{m}(\kappa r_{\mathrm{g}})}{(J_{m-1}(k r_{\mathrm{g}})-J_{m+1}(k r_{\mathrm{g}})) Y_m(k r_{\mathrm{g}})+J_m(k r_{\mathrm{g}}) (Y_{m+1}(k r_{\mathrm{g}})-Y_{m-1}(k r_{\mathrm{g}}))}.
\end{aligned}
\end{equation}
\end{widetext}
The corresponding phase shifts have been used to calculate the scattering observables, which are presented in the main text of the paper. 
\section{Perturbative scattering theory} \label{SuppMatt_3}
This section provides details on the derivation of Eq.~(4) in the main part of the Letter. The differential cross-section for the plasma wave scattering
\begin{equation}
\frac{d\sigma}{d\phi}=\frac{k}{2 \pi u^2} \cdot |T_{\vec{k}'\vec{k}} (\Omega_\vec{k})|^2. 
\end{equation}
is determined by the T-matrix, which in the second-order Born approximation is approximated as $\hat{T}_{\vec{k}' 
\vec{k}}(\omega)=\hat{V}_{\vec{k}' 
    \vec{k}} + \sum_{\vec{k}''}\hat{V}_{\vec{k}' 
    \vec{k}''}\hat{G}(\omega, \vec{k}'')\hat{V}_{\vec{k}'' 
    \vec{k}}$. Here $G(\omega,\vec{k})=(\omega-\hat{H}(\vec{k})+i \epsilon)^{-1}$ is the Green function for the effective Hamiltonian $\hat{H}(\vec{k})$ describing the plasma waves. Up to third order in the potential $\hat{V}$, the differential cross section is given by
\begin{equation*}
\begin{split}
    \frac{d \sigma}{d \phi} \approx  \frac{k}{2 \pi u^{2}} \Big(|\hat{V}_{\mathbf{k}^{\prime} \mathbf{k}}|^{2} + 2 \sum_{k''} \Re\big[\hat{V}_{\vec{k}' \vec{k}''} \hat{V}_{\vec{k}'' \vec{k}} \hat{V}_{\vec{k} \vec{k}'} \big] \times \\ G'(\Omega_\vec{k},\vec{k}'') -  2 \sum_{\vec{k}''} \Im\big[\hat{V}_{\vec{k}' \vec{k}''} \hat{V}_{\vec{k}'' \vec{k}} \hat{V}_{\vec{k} \vec{k}'} \big]G''(\Omega_\vec{k},\vec{k}'')\Big). 
 \end{split}
\end{equation*}
Here $G'(\omega,\vec{k})$ and $G''(\omega,\vec{k})$ are real and imaginary parts of the Green function. The explicit calulation of the matrix element $\hat{V}_{\vec{k} \vec{k}'}=i\pi r_{\mathrm{g}} \omega_{\mathrm{g}}J_{1}( k r_{\mathrm{g}})\sin(\phi_\vec{k}-\phi_\vec{k}')/k$ demonstrates that it has only the imaginary part and the second term in the equation above vanishes. If we incorporate the explicit expression for the imaginary part of the Green function $G''(\Omega_\vec{k},\vec{k}'')=-\pi \delta (\Omega_{\vec{k}''} - \Omega_{\vec{k}})$, the resulting expression for the cross-section matches with Eq.~(4) presented in the main part of the Letter. 
\onecolumngrid
\begin{center}
\begin{figure}[t]
\centering
    \includegraphics[trim=3cm 3.5cm 4.5cm 2cm, width=1.\columnwidth]{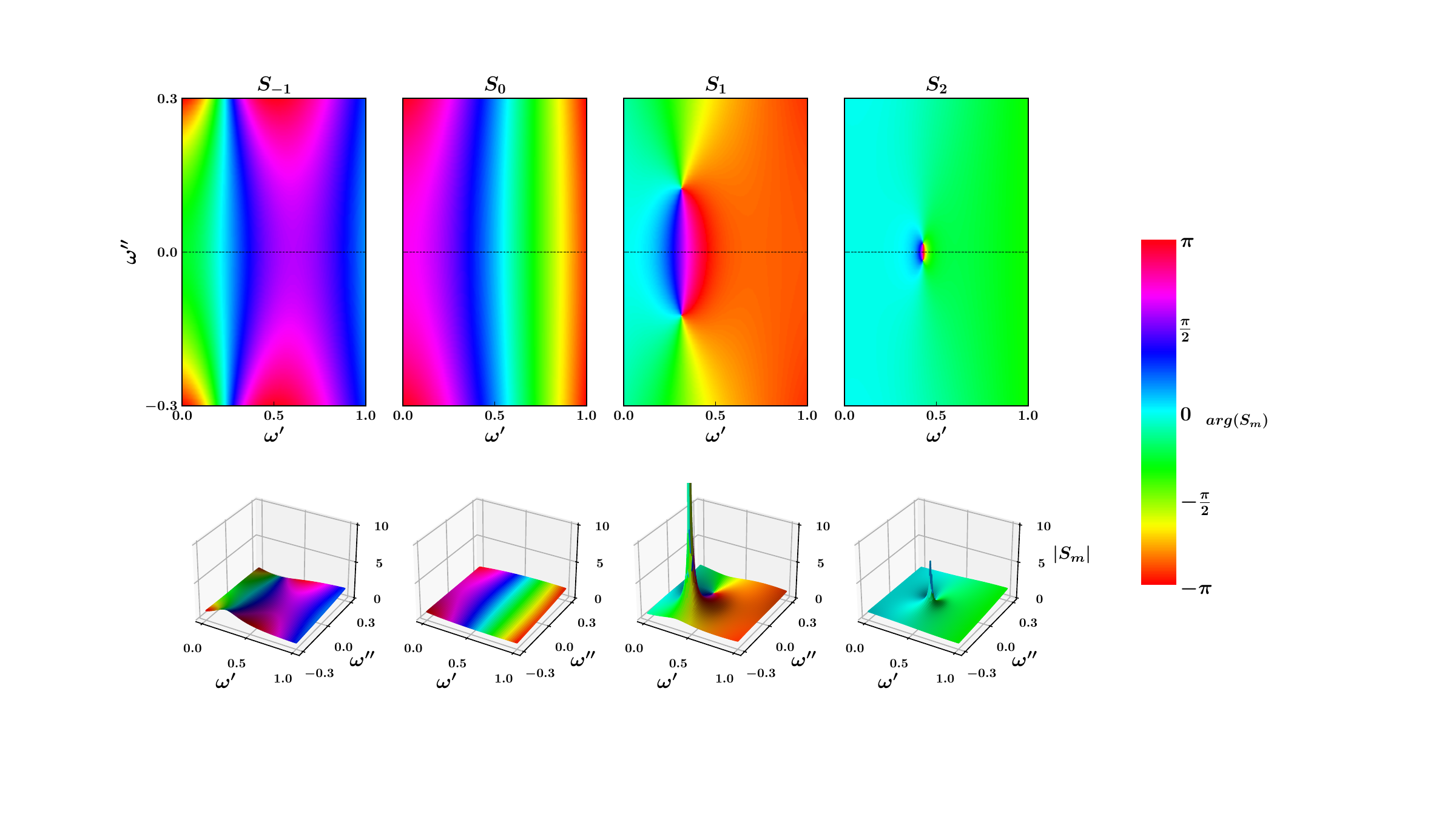}
    \caption{S-matrix plotted in the complex plane for harmonics $m=-1, 0, 1, 2$. Shown in the top row are complex phase plots for each S-matrix harmonic, as labeled, and in the bottom row we present the same data but with the magnitude of the complex S-matrix projected on the z-axis.}
    \label{SuppFig1}
\end{figure}
\end{center}
\twocolumngrid

\section{Poles of the S-matrix}
The resonant nature of the scattering can be tracked in the frequency behavior of the S-matrix $S_{m}(\omega)=e^{2 i \delta_{m}(\omega/\omega_{\mathrm{g}})}$ extended to the complex plane as $\omega=\omega'+i \omega''$. The corresponding frequency dependence for orbital numbers $m=-1,0,1,2$ is presented in Fig.~(\ref{SuppFig1}). The resonances manifest as poles in the lower part of the complex plane.   Notably, a single pole for partial waves with $m\geq 1$ exists. In contrast, the partial waves with $m\leq 0$ are free of any poles in the complex plane. This behavior demonstrates that the presence of the resonant plasma wave scattering is due to the trapped chiral mode circulating the micromagnet.   
\section{Extra metallic screening and gating by the micromagnet}
\begin{figure}[ht]
    \centering
    \vspace{-0.2cm}
    \includegraphics[clip,trim={0cm 0cm 0cm 0cm}, width=\columnwidth]{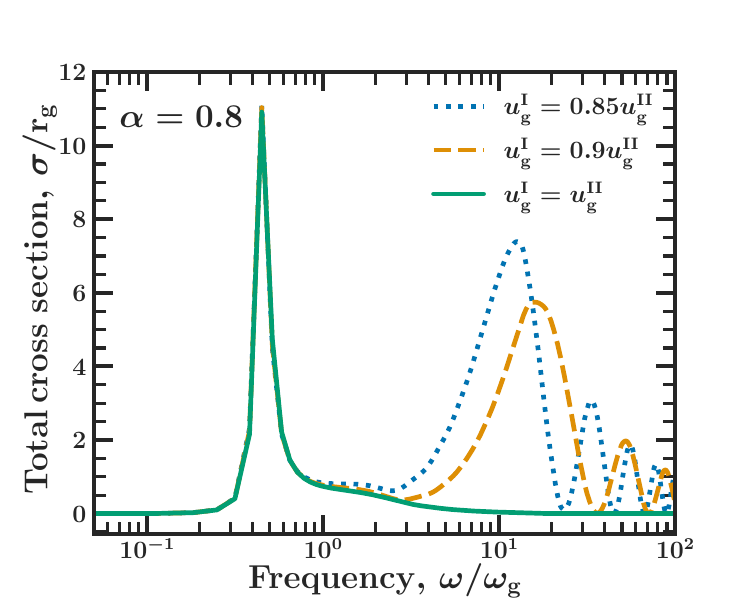}
    \caption{Frequency dependence of total cross section, $\sigma$, when the plasma wave velocity is equal across interface (green) and when the velocity is altered mismatched by $u_{\mathrm{g}}^{\mathrm{I}}=0.85 u_{\mathrm{g}}^{\mathrm{II}}$ (blue, dotted) and $u_{\mathrm{g}}^{\mathrm{I}}=0.9u_{\mathrm{g}}^{\mathrm{II}}$ (orange, dashed). Calculated for scattering strength $\alpha$ as labeled.}
    \label{Non-uniform calculations}
    \vspace{-0.3cm}
\end{figure}
The micrometer-sized magnets Ni, NiFe, and NiCr used in similar setups~\cite{Micromagnets1,Micromagnets2,Micromagnets3,Micromagnets4,Micromagnets5} are metallic and provide additional screening and gating. They modify the plasma wave velocity below the ferromagnet  as $u_\mathrm{g}=u\sqrt{n_\mathrm{g} d_\mathrm{g}/ n (d+d_\mathrm{g})}$. Here $d_\mathrm{g}$ is the distance to the micromagnet, and $n_\mathrm{g}$ is electron concentration below it. Since the velocity of plasma waves manifests itself explicitly in the expression for the spinor $\psi (\mathbf{r},t) =\left[j_{+}(\mathbf{r},t), u \rho (\mathbf{r},t),j_{-}(\mathbf{r},t)\right]$ the boundary condition which ensures the continuity of the electric potential needs to be modified as follows 
\begin{equation}
\frac{1}{u_{\mathrm{g}}^\mathrm{I}}\psi_{m}^{0, \mathrm{I} }(r_{\mathrm{g}})  = \frac{1}{u_{\mathrm{g}}^\mathrm{II}}\psi_{m}^{0, \mathrm{II}}(r_{\mathrm{g}}),
\end{equation}
Here $u_{\mathrm{g}}^{\mathrm{I} (\mathrm{II})}$ are plasma wave velocities at both sides of the interface. Otherwise, the calculation of the phase shifts $\delta_m(\vec{k})$ follows the same steps, which are presented in Sec.~\ref{secII}. 

For calculations, we chose the spacing between graphene and the micromagnet as $d_{\mathrm{g}}=50\;\hbox{nm}$ and $d_{\mathrm{g}}=30\;\hbox{nm}$. As a result, plasma wave velocity below the micromagnet is reduced as $u_{\mathrm{g}}^{\mathrm{I}}\approx 0.9 u_{\mathrm{g}}^{\mathrm{II}}$ and $u_{\mathrm{g}}^{\mathrm{I}}\approx 0.85 u_{\mathrm{g}}^{\mathrm{II}}$, respectively. For these estimations, we also assume that the change of electron density across the interface is negligible, i.e., $n_\mathrm{g}\approx n$. The corresponding frequency dependence of the total cross-section for the plasma waves scattering (supplemented by the curve for $u_{\mathrm{g}}^{\mathrm{II}}=u_{\mathrm{g}}^{\mathrm{I}}$, which corresponds to the absence of the additional screening and gating) is presented in Fig.~(\ref{Non-uniform calculations}). The oscillations we see at high frequencies in Fig.~(\ref{Non-uniform calculations}) are due solely to the mismatch of the plasma wave velocity, and the scattering in this frequency range has a negligible skewness.  Furthermore, the low-frequency resonance is extremely robust to the velocity mismatch. We conclude that the possible velocity mismatch across the magnetic field interface minimally affects the giant resonant skew scattering of plasma waves predicted in the Letter. 
\onecolumngrid
\begin{center}
\begin{figure}[t]
\centering
        \includegraphics[trim=3cm 0.8cm 2.cm 1.2cm, clip,width=1\textwidth]{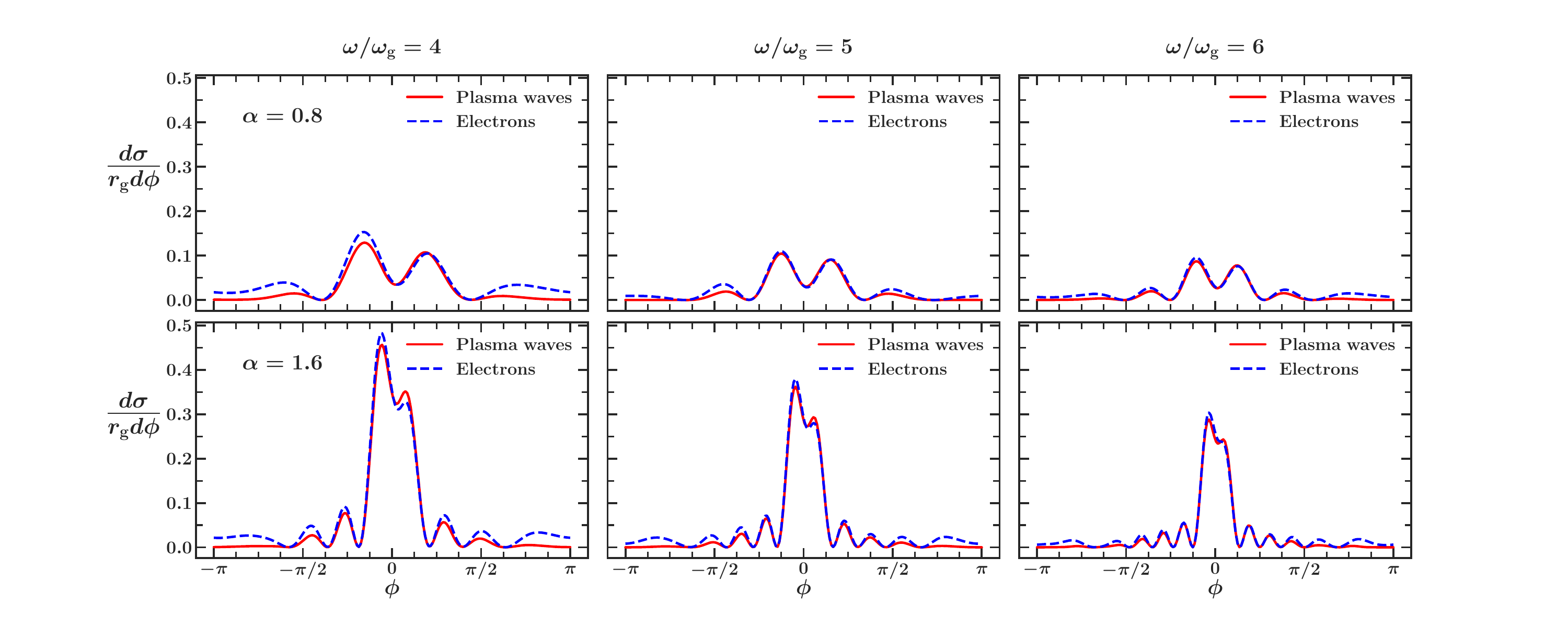}
        \caption{Angular dependence of the differential cross section $d\sigma / d\phi$ for high frequency cuts $\omega / \omega_{\mathrm{g}}$ as labelled and scattering strengths $\alpha$ as labelled.}
        \label{Round 1 Fig. 2}
\end{figure}
\end{center}
\twocolumngrid
\section{The high-frequency behavior for electron and plasma wave scattering} \label{Section VI}
In the main text of the Letter, we showed that the low-frequency ($\omega \lesssim \omega_{\mathrm{g}}$) scattering of plasma waves and Dirac electrons are drastically different. 
Fig.~(\ref{Round 1 Fig. 2}) presents three frequency cuts ($\omega/\omega_\mathrm{g}=4,\; 5,\; 6$) corresponding to the differential cross-section in the high-frequency regime. In this regime, not only is the forward scattering ($|\phi|\lesssim\pi/4$) dominant, but its angular profile is almost indistinguishable for plasma waves and Dirac electrons. The deviations become apparent only if the scattering angle is large enough, $|\phi|\gtrsim \pi/2$.

The similarity of forward scattering for plasma waves and Dirac electrons does not rely on the circular symmetry of the target. This can be confirmed using the  perturbative expression for the differential cross-section, Eq.~(4) from the main text, and a target with the elliptical shape, i.e. $\omega_{c}(x,y) = \omega_{\mathrm{g}}\Theta\left(1-(x/r_\mathrm{g}^x)^2 -  (x/r_\mathrm{g}^y)^2\right)$. Its Fourier transform can be calculated analytically and is given by $ \omega_{c}(k_{x},k_{y})= 2 \pi \omega_{g} r_\mathrm{g}^x  r_\mathrm{g}^y J_{1}\left[\Lambda(k_{x},k_{y})\right]/\Lambda(k_{x},k_{y})$,  where $\Lambda(k_{x},k_{y})=\sqrt{(r_\mathrm{g}^x k_{x})^{2}+(r_\mathrm{g}^y k_{y})^{2}}$. The angular distribution of the differential cross-section is presented in Fig.~(\ref{Round 1 elliptic}). The two columns correspond to the different ellipse orientations with respect to the incoming plasma or electronic wave ($r_\mathrm{g}^x/r_\mathrm{g}^y \approx 0.7$ and $r_\mathrm{g}^x/r_\mathrm{g}^y \approx 1.4$). It is clearly seen that the angular distribution of differential cross-section in the forward-scattering sector is still almost indistinguishable for the two models.

\onecolumngrid
\begin{center}
    \begin{figure}[t]
    \vspace{-1cm}
    \includegraphics[width=0.8\textwidth]{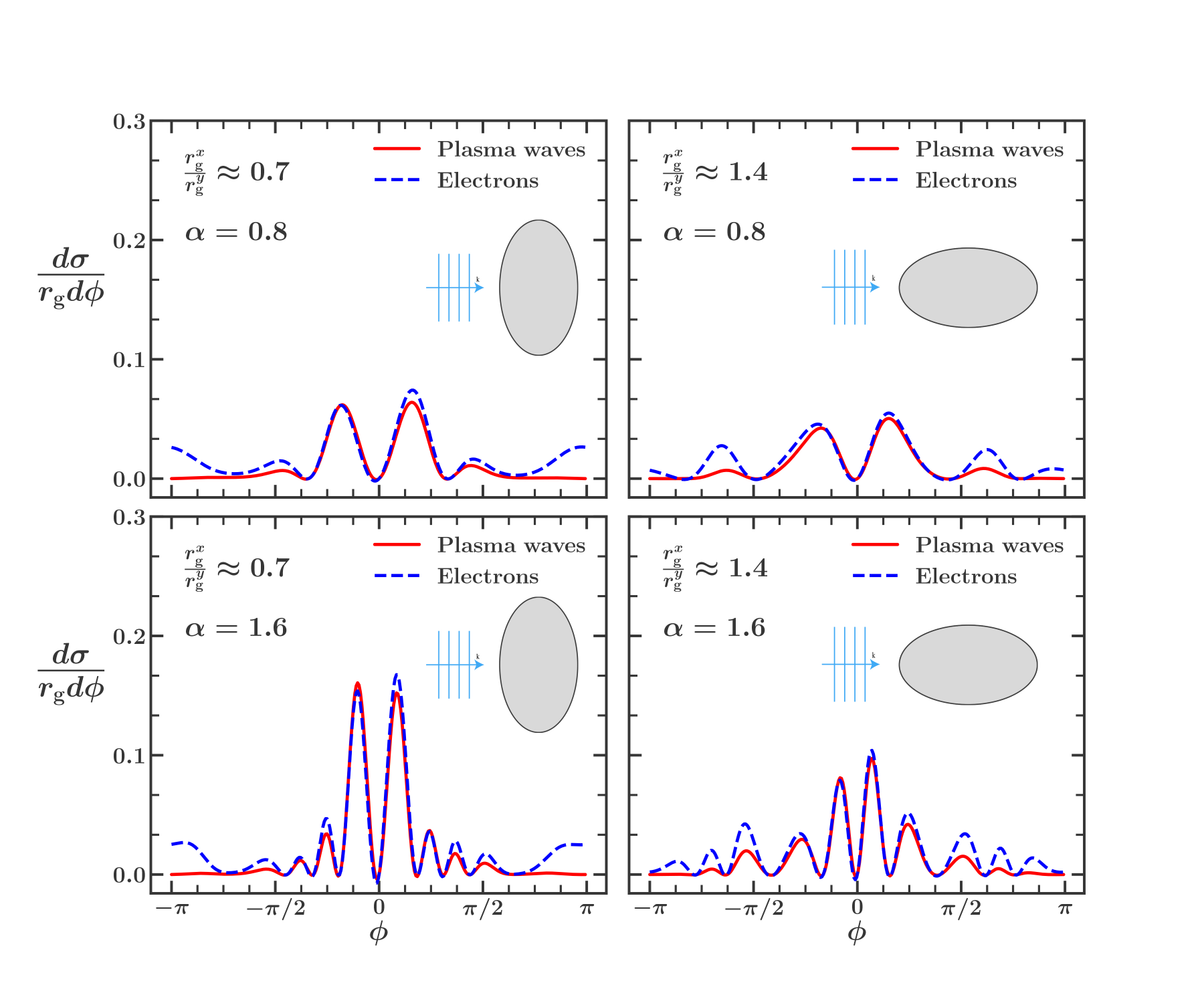}
    \vspace{-0.5cm}
    \caption{Angular dependence of differential cross section $d\sigma/d\phi$ for a spatially elliptic scattering profile. Parameters for the shape of the ellipse are as labelled and insets show the orientation of the scatterer in real space. Scattering strengths $\alpha$ are as labelled. All curves are produced for an incident frequency $\omega/\omega_{\mathrm{g}}=5$.}
    \label{Round 1 elliptic}
    \end{figure}
    \vspace{-0.9cm}
\end{center}
\twocolumngrid

\section{The magnetic field profile of the ferromagnetic column}
The considered step-like profile of the magnetic field, $B_z=B_{\mathrm{g}} \Theta (r_\mathrm{g}-r)$, corresponds to the ferromagnetic column with radius $r_\mathrm{g}$ and an infinite height. Here, we argue that the main results of our Letter are valid even if the column height $h_\mathrm{g}$ is comparable with its radius.

According to the magnetostatics, it is convenient to solve the system of Maxwell equations ($\hbox{div} \ \vec{B}(\vec{r})=0$ and $\hbox{curl} \ \vec{H}(\vec{r})=0$) supplemented by the matter response relation ($\vec{B}(\vec{r})=\vec{H}(\vec{r})+4\pi \vec{M}(\vec{r})$) using the scalar magnetic potential $\Phi_\mathrm{m}(\vec{r})$. The potential determines the magnetic field as $\vec{H}(\vec{r})=-\nabla \Phi_\mathrm{m}(\vec{r})$ and satisfies the Laplace equation $\Delta \Phi_{m}(\mathbf{r}) = -4 \pi \rho_{m}(\mathbf{r})$, where $\rho_{m}(\vec{r})=-\mathrm{div}\vec{M}(\vec{r})$ is usually interpreted as the effective magnetic charge density. 

\begin{figure}[h]
    \centering \includegraphics[width=1\columnwidth]{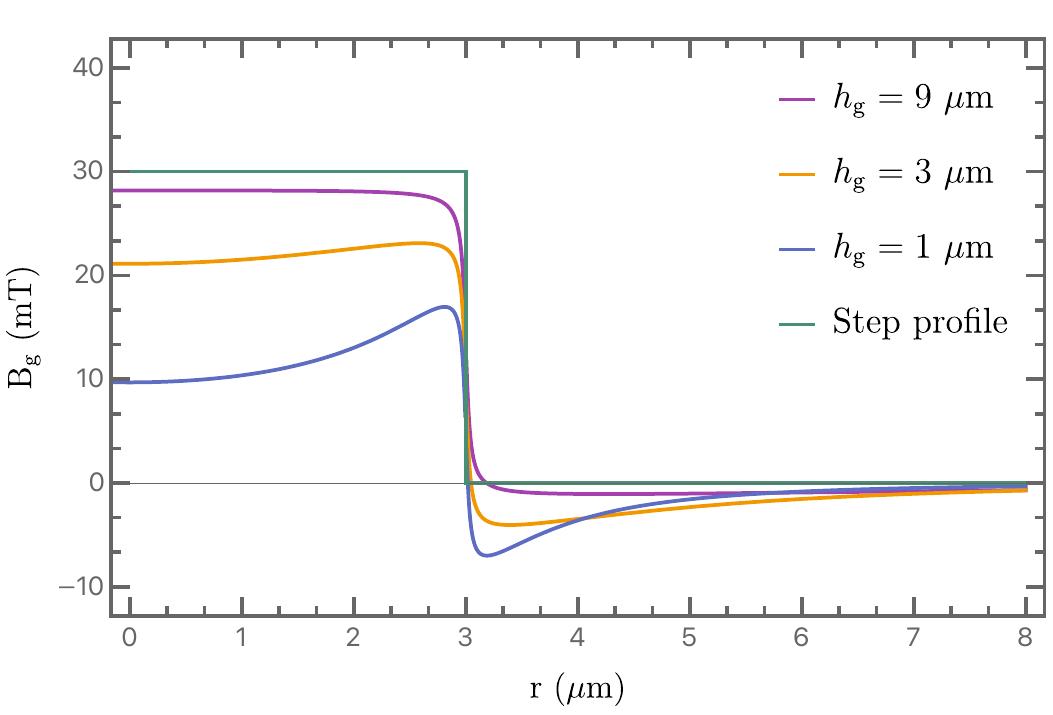}
    \caption{Magnetic field profile of column micromagnets with lengths $\mathrm{h}_{\mathrm{g}}$ as labeled, plotted with the step profile for comparison. All curves were calculated with a micromagnet radius $r_{g}=3 \ \mu \mathrm{m}$, distance between magnet and electron gas of $d=30 \ \mathrm{nm}$, and magnetisation $M_{0}=30/2\pi \ \mathrm{mT}$.}
    \label{FigMagneticFieldProfile}
\end{figure}

If the ferromagnetic column is placed at a distance $d_\mathrm{g}$ from the graphene monolayer, the distribution of the magnetic charge can be presented as
\begin{equation*}
    \rho_{m}(\vec{r}) = M_{0}\left[\delta(z-d_\mathrm{g}) - \delta(z-d_\mathrm{g} -h_{\mathrm{g}}) \right]\theta(r_{\mathrm{g}}-r), \label{Round 1 eq.2}
\end{equation*}
Here $M_0$ is the magnetization inside the ferromagnet. The magnetic charge distribution is similar to the distribution of electric charge inside a capacitor with circular plates. The explicit magnetostatic calculation results in the following magnetic field distribution 
\begin{equation*}
    B_{z}(r) = -2 \pi M_{0} \left[F\left(\frac{r}{r_{\mathrm{g}}},\frac{d_\mathrm{g}}{r_{\mathrm{g}}}\right) - F\left(\frac{r}{r_{\mathrm{g}}}, \frac{d_\mathrm{g}+h_\mathrm{g}}{r_{\mathrm{g}}}\right) \right]. \label{Round 1 eq.3}    
\end{equation*}
It is shaped by the function $F(r,d)$, which is given by 
\begin{equation}
    F(r,d) = \int_{0}^{\infty} d\gamma J_{0}(\gamma) J_{1}(\gamma r)e^{-\gamma d}. \label{Round 1 eq.4}
\end{equation}
The numerical calculations of the magnetic field profile are presented in Fig.~(\ref{FigMagneticFieldProfile}). We considered three values of the column height: $h_\mathrm{g}=9 \ \mu\hbox{m}$ (purple), $3 \ \mu\hbox{m}$ (orange) and $1 \ \mu\hbox{m}$ (blue). We also assume a separation between the graphene layer and ferromagnetic column of $d_{\mathrm{g}}=30 \ \mathrm{nm}$ and radius of $r_\mathrm{g}=3\; \mu \hbox{m}$. The step-like function can reasonably approximate the magnetic field profile even when the column height is only slightly larger than its radius. The deviations become apparent when the shape of the ferromagnet becomes closer to a tablet rather than a column. 

\begin{figure}[t]
    \centering
    \vspace{-0.1cm}
    \includegraphics[width = \columnwidth, height=110mm]{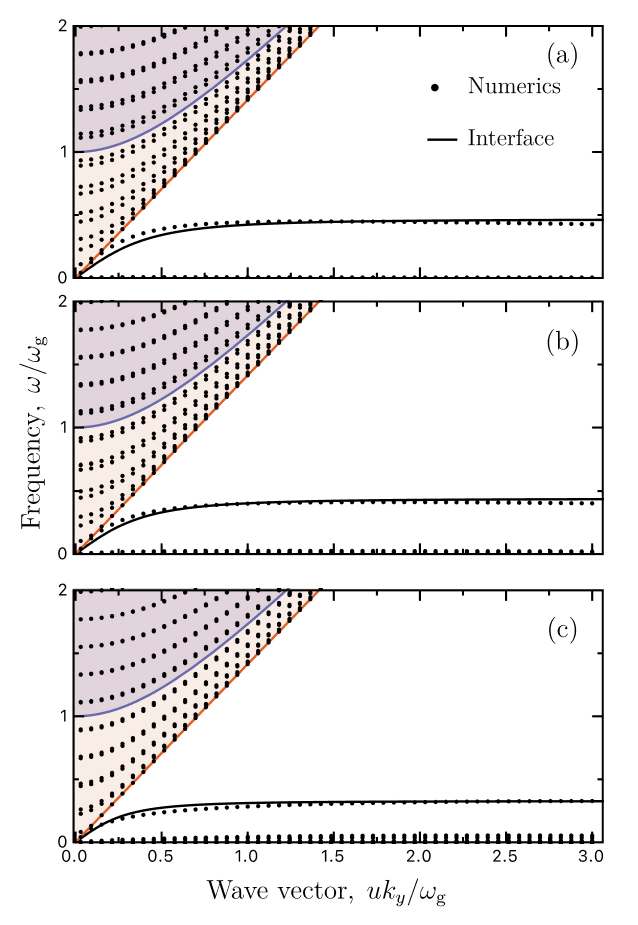}
    \caption{Spectrum of plasma waves in the presence of a column magnetic field profile calculated numerically (black dots), plotted with the analytic expression for the interface mode (black). We consider column ferromagnets of length (a): $9 \ \mu \mathrm{m}$, (b): $3 \ \mu \mathrm{m}$, and (c): $1 \ \mu\mathrm{m}$. Superimposed on all panels are the bulk continuum of states outside (orange) and inside (blue) the region under the magnetic field. }
    \label{Round 1 Fig. 4}
\end{figure}

We considered the finite height effects on the chiral interface plasma wave circulating the ferromagnet. Its presence and chiral nature are responsible for giant resonant skew scattering of plasma waves off the micromagnet.  It is instructive to shift to the 1D interface geometry we considered in the main letter, with a cyclotron frequency profile $\omega_\mathrm{c}(x)=e B_z(x)/mc$ following the radial distributions of magnetic field profiles $B_z(r)$ calculated above in Fig.~(\ref{FigMagneticFieldProfile}). This problem can be straightforwardly discretized. The spectrum of the interface mode has been evaluated numerically and is presented in Fig.~(\ref{Round 1 Fig. 4}) for column magnet heights $h_{\mathrm{g}}=9 \ \mu\mathrm{m}$, $3\ \mu\mathrm{m}$, and $1 \ \mu\mathrm{m}$. It is instructive to compare them with the analytically evaluated dispersion of the interface mode for the step-like profile $B_z(x)=\bar{B} \theta (r_\mathrm{g}-x)$ with a phenomenologically introduced magnetic field jump $\bar{B}$ across the interface. A good fit is achieved if the latter is chosen as $\bar{B} = \int_{0}^{r_{\mathrm{g}}}B_{z}(r) dr/r_\mathrm{g} -B_{z}^{\mathrm{min}}$, where $B_{z}^{\mathrm{min}}$ is the minimum magnetic field strength (corresponding to the troughs in Fig.~(\ref{FigMagneticFieldProfile})) and the first term is the average magnetic field under the column ferromagnet.  Thus the chiral interface wave can be accurately described by the step-like profile of magnetic field and the giant resonant scattering of plasma waves is robust to the finite height effects of the ferromagnetic column.    

\end{document}